\documentclass[doublecol]{epl2}
\usepackage{graphics,bm}
\usepackage{amssymb}
\usepackage{epsfig}
\usepackage{epsf}
\def\Re{\rm{Re}}
\def\Im{\rm{Im}}
\def\be{\begin{equation}} \def\ee{\end{equation}}
\def\bea{\begin{eqnarray}} \def\eea{\end{eqnarray}}

\def\nn{\nonumber}

\title{Orbital Resonance Mode in Superconducting Iron Pnictides}

\author{Wei-Cheng Lee\inst{1} \and Philip W. Phillips\inst{1}}
\shortauthor{W.-C. Lee \etal}

\institute{
  \inst{1} Department of Physics, University of Illinois, 1110 West Greent Stree, Urbana, Illinois 61801, USA
}


\abstract{
We show that the fluctuations associated with ferro orbital order in
the $d_{xz}$ and $d_{yz}$ orbitals can develop a sharp resonance mode
in the superconducting state with 
a nodeless gap on the Fermi surface. 
This orbital resonance mode appears below the particle-hole continuum
and is analogous to the magnetic resonance mode found 
in various unconventional superconductors. If the pairing symmetry is $s_{\pm}$, a dynamical coupling between the orbital ordering and the $d$-wave subdominant pairing channels is 
present by symmetry. Therefore the nature of the resonance mode depends on the relative strengths of the fluctuations in these two channels, which could vary significantly 
for different families of the iron based superconductors. The
application of our theory to a recent observation of a new
$\delta$-function-like peak in the B$_{1g}$ Raman spectrum
of Ba$_{0.6}$K$_{0.4}$Fe$_2$As$_2$ is discussed, and we predict that the same orbital resonance mode can be detected in electron-energy-loss-spectroscopy (EELS).
}

\begin{document}

\maketitle
\section{Introduction}
For high-temperature superconductors, cuprates and iron pnictides, resolving the nature of the fluctuations in both the
normal and superconducting states remains a crucial question as it holds the key to the pairing mechanism.
In cuprates, due to the strong antiferromagnetism in the parent compounds, it is widely-believed that the antiferromagnetic spin fluctuations are the most important 
ingredient in the pairing mechanism. 
One of the marquee indicators of this is the magnetic resonance mode\cite{scalapino2012} 
observed in the superconducting state of every cuprate. 
From the BCS theory, the spin-flip susceptibility of an electron scattered from $\vec{k}$ to
$\vec{k}+\vec{q}$ in the superconducting state gains an extra coherence factor which is proportional to $(1-{\rm sgn}(\Delta(\vec{k})){\rm sgn}(\Delta(\vec{k}+\vec{q})))$. 
Since the gap symmetry of the cuprates is $d$-wave signified by ${\rm sgn}(\Delta(\vec{k})) = -{\rm sgn}(\Delta(\vec{k}+\vec{Q}))$, 
the spin excitations near $\vec{q}=\vec{Q}$ are compatible with superconductivity.
It can be further shown\cite{demler1998, brinckmann1999,tchernyshyov2001,leewcinsr1,leewcinsr2} that a sharp $\delta$-function-like resonance mode 
in the spin-flip susceptibility requires an antiferromagnetic spin
interaction to pull the resonance mode to an energy below the particle-hole continuum.
In other words, despite some dependence on the detailed electronic
structure, the existence of the magnetic resonance mode is
predominantly determined by the gap symmetry and
the nature of the spin interaction.
As a result, it has been identified as an unambiguous\cite{scalapino2012} indicator that
antiferromagnetic spin fluctuations remain strong in the superconducting state and consequently might be the
driving mechanism for superconductivity.

In iron pnictides, the structural phase transition followed by 
stripe-like antiferromagnetism is a robust feature in the phase diagram. 
The onset of the structural phase transition breaks the $C_4$ symmetry
down to $C_2$ which enables magnetic order at wavevector $(\pi,0)$ 
and orbital order at zero wavevector characterized by unequal occupation in $d_{xz}$ and $d_{yz}$ orbitals.
These two unique consequences suggest that the
dominant fluctuations might be either orbital-based\cite{lv2009,kruger2009,leewc2009,lee_cc2009,chen_cc2010,lv2010,kontani2011} or spin-based 
\cite{yildirim2008,xu2008,chen_f2008,fernandes2010,fernandes2012}.
Indeed, a magnetic resonance mode at wavevector $(\pi,0)$ has been
observed in the pnictides and has been attributed to spin
fluctuations\cite{scalapino2012}.  However, recent Raman scattering
measurements\cite{kretzschmar2012}, a zero wave-vector probe, have
observed a superconductivity-induced peak in the $B_{1g}$ channel. 
Since this peak is sharp and occurs at zero wavevector, for two reasons, it is not
likely that it is due to
to spin fluctuations. First, because spin-nematicity corresponds to
 a breaking of the $Z_2$ symmetry between spin fluctuations at 
$(\pi,0)$ and $(0,\pi)$, the associated fluctuations, if any, reside
in the four-spin susceptibility 
$\langle \hat{T}_t(\hat{S}^2_x(t)-\hat{S}_y^2(t))(\hat{S}^2_x(0)-\hat{S}_y^2(0))\rangle$\cite{fernandes2012}. 
Such a high-order spin susceptibility couples to the Raman vertex via complicated matrix elements which cannot produce a sharp delta-function-like peak in Raman spectroscopy. 
Moreover, since the Raman peak in question is observed in superconducting samples which are doped away from the parent compounds, 
the flucutations associated with the $Z_2$ spin nematicity are likely to diminish, and only the spin fluctuations around $(\pi,0)$ and $(0,\pi)$ survive.
It is indeed true that Raman spectroscopy can detect large-momentum spin excitations through two-magnon processes, but this usually gives rise to 
broad peaks\cite{ramanreview}. In contrast, since orbital fluctuations
reside in the $B_{1g}$ charge channel 
a resonance mode develops due to the orbital fluctuations which can certainly be detected by the Raman spectroscopy.
Such a resonance mode can, in principle, arise from a
sub-dominant $d$-wave pairing channel\cite{leewc2009trs,scalapino2009,kretzschmar2012}.  
Our key point here is that  $d$-wave sub-dominant pairing and
ferro-orbital order both have the same space group symmetry of the
Raman $B_{1g}$ mode.  Consequently, the observed
 $B_{1g}$ mode in Raman could arise from orbital fluctuations.  We
 term this mode an {\it orbital resonance mode}.
We work out the details of this scenario and show that the observed $B_{1g}$
mode could offer an unprecedented fingerprint of ferro-orbital order in the
pnictides.  The Raman experiments then supplement the current indirect
evidence in the non-superconducting states that point-contact spectroscopy\cite{arham2011,arham2012,leewcnfl2012} 
and neutron scattering measurements\cite{xu2012,leewcincomm2012}
provide for orbital fluctuations.

\section{Criteria for the orbital resonance mode}
First we give a general discussion for the existence of the orbital
resonance mode as was done for the magnetic resonance mode. 
Given that orbital fluctuations are in the charge channel, the extra coherence factor from BCS theory now becomes 
$\sim (1+{\rm sgn}(\Delta(\vec{k})){\rm sgn}(\Delta(\vec{k}+\vec{q})))$. 
We immediately see that the sharp difference from the magnetic
counterpart is that the pre-requisite condition for the orbital resonance mode 
is ${\rm sgn}(\Delta(\vec{k})) = {\rm sgn}(\Delta(\vec{k}+\vec{q}))$. 
Since we focus on the case with wavevectors near zero ($\vec{q}\sim 0$), such a condition generally holds in any gap symmetry. 
In the following, we show that the orbital resonance mode generally
exists provided an effective attractive interaction is present in the
orbital ordering channel.  

Before moving to the microscopic calculations, we prove that the orbital ordering and $d$-wave sub-dominant pairing channels are coupled in an
$s$-wave superconducting state. 
This state of affairs obtains for two crucial reasons.
First, because the singlet Cooper pair is a mixture of electrons and holes with different spins, the particle-hole excitations are intrinsically coupled to 
the particle-particle excitations as long as the space group symmetry allows.
Second, both orbital ordering and the $d$-wave pairing channels have a sign change
under a rotation of $\pi/2$ but the $s$-wave ground state does not.
Therefore a Berry-phase coupling between them is allowed by symmetry
as discussed in Ref. [\cite{leewc2009trs}].  This is the key
physical principle underlying our work here, and we will show that such a Berry phase coupling enriches the physics of the $B_{1g}$ resonance mode.

\section{Formalism}
We start from a general two-orbital model of the superconducting state for iron pnictides.  The model Hamiltonian is given by
\bea
H&=&H_0 + H_{SC} + H_I,\nn\\
H_0 &=& \sum_{\vec{k},\sigma} \psi^\dagger_{\vec{k},\sigma}\left[\epsilon_+(\vec{k})\hat{I} + \epsilon_-(\vec{k})\hat{\tau}_3 + \epsilon_{xy}(\vec{k})\hat{\tau}_x\right]
\psi_{\vec{k},\sigma}\nn\\
&\equiv& \sum_{\vec{k},\sigma} \phi^\dagger_{\vec{k},\sigma}\hat{D}_{\vec{k},\sigma}\phi_{\vec{k},\sigma},\nn\\
H_{SC}&=& \sum_{\vec{k}} \Delta(\vec{k})\left[ \alpha^\dagger_{-\vec{k},\downarrow} \alpha^\dagger_{\vec{k},\uparrow} + 
\beta^\dagger_{-\vec{k},\downarrow} \beta^\dagger_{\vec{k},\uparrow} + h.c.\right],
\eea
where $\psi^\dagger_{\vec{k},\sigma} \equiv (d^\dagger_{\vec{k},xz,\sigma}, d^\dagger_{\vec{k},yz,\sigma})$ are the electron creation operators for 
orbital $xz$ or $yz$ with spin $\sigma$, $\hat{D}_{\vec{k},\sigma} = 
diag. (E^{\alpha}_{\vec{k},\sigma},E^{\beta}_{\vec{k},\sigma})$ the eigenenergy, and
$\phi^\dagger_{\vec{k},\sigma} \equiv (\alpha^\dagger_{\vec{k},\sigma}, \beta^\dagger_{\vec{k},\sigma})$ are the corresponding eigenvectors.
We adopt the model proposed by Raghu {\it et. al.}\cite{raghu2008} for $H_0$ and use the same tight-binding parameters.
$H_{SC}$ describes the mean-field pairing potential in the diagonalized basis and we assume the pairing symmetry to be $s_{\pm}$, which in the two-orbital model is
$\Delta(\vec{k}) = \Delta_0\,\cos(kx)\cdot\cos(ky)$.
Without loss of generality, we consider two types of interactions $H_I = H_{OO} + H_{dSC}$, 
\bea
H_{OO} &=& \frac{1}{N}\sum_{\vec{q}} V_{OO}({\vec{q}})\hat{O}(-\vec{q})\hat{O}(\vec{q}),\nn\\
H_{dSC} &=& \frac{1}{N}\sum_{\vec{q}} V_d(\vec{q}) \left[\Delta^d(\vec{q})\right]^\dagger \Delta^d(\vec{q}),
\eea
where 
\bea
\hat{O}(\vec{q}) &=& \sum_{\vec{k},\sigma} \psi^\dagger_{\vec{k}+\vec{q},\sigma}\hat{\tau}_3\psi_{\vec{k},\sigma},\nn\\
\Delta^d(\vec{q}) &=& \sum_{\vec{k}} d_{\vec{k}} \left[\alpha_{\vec{k},\uparrow}\alpha_{-\vec{k}+\vec{q},\downarrow} + 
\beta_{\vec{k},\uparrow}\beta_{-\vec{k}+\vec{q},\downarrow}\right],\nn\\
d_{\vec{k}} &\equiv& \frac{\cos k_x -\cos k_y}{2}.
\label{orderparameter}
\eea
Orbital order is then characterized by $\langle \hat{O}(\vec{q}=0)\rangle\neq 0$, 
and $H_{OO}$ is the effective interaction in the orbital ordering channel. The orbital ordering instability 
requires $V_{OO}(\vec{q}=0)$ to be negative and smaller than a critical value.
Finally, $H_{dSC}$ represents the interaction for the sub-dominant
pairing interaction in the $d$-wave channel, which is generally believed\cite{kuroki2008,wangfa2009,seo2009,graser2009} to exist in iron-based superconductors due to their 
unique Fermi surface topology. 
Because we take the ground state to be the $s_{\pm}$ superconducting state with which the $d$-wave pairing channel is competing, the instability toward $d$-wave superconductivity will occur only if $V_d(\vec{q}=0)$ exceeds a critical value.
In other words, the ground state of $s_{\pm}$ superconductivity is stable as long as $V_{OO}(\vec{q}=0) > - V^c_{OO}$ and $V_d(\vec{q}=0) > -V^c_d$, where $V^c_{OO},V^c_d > 0$.
Note that both terms can be obtained from a microscopic multi-orbital Hubbard Hamiltonian.

\begin{figure}
\includegraphics{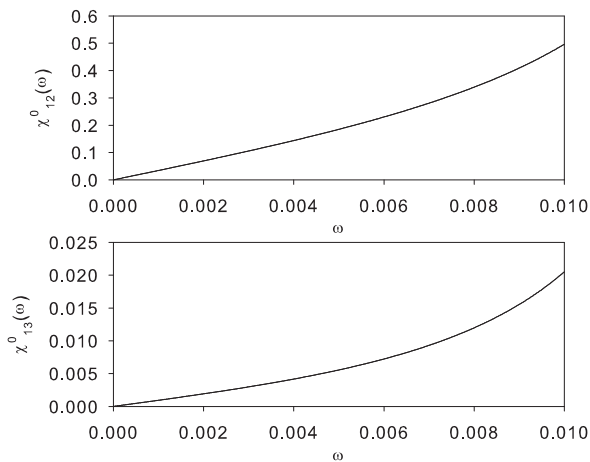}
\caption{\label{fig:coupling} The dynamical couplings between the orbital ordering and $d$-wave pairing channels. 
The magnitude of the gap is $\Delta_0 = 0.01$, and the edge of the particle-hole continuum $\Omega_{ph}$ is 0.0127. 
A typical behavior of the Berry phase coupling: $\chi(\omega)\sim \omega$ can be seen.}
\end{figure}

Since our main purpose is to study the behavior of the fluctuations due to $H_{OO}$ and $H_{dSC}$ in a stable $s_{\pm}$ superconductor, we choose $V_{OO}(\vec{q}=0)$ and 
$V_d(\vec{q}=0)$ to be close but not exceeding critical values. The
procedure for analyzing our model Hamiltonian is well established.
First, the Bogoliubov transformation is performed to diagonalize $(H_0
+ H_{SC})$.  Second, the random-phase approximation is applied to the
two-particle correlation functions to express them in terms of the 
Bogoliubov quasiparticles. The relevant susceptibility is given by 
\be
\big(\hat{\chi}(\vec{q},\omega)\big)^{-1} = \big(\hat{\chi}^0(\vec{q},\omega)\big)^{-1} - \hat{U}_{\vec{q}}
\ee
$\hat{\chi}^0(\vec{q},\omega), \hat{\chi}(\vec{q},\omega)$, and
$\hat{U}_{\vec{q}}$ are all $3\times 3$ matrices.  The bare
correlation functions, $\hat{\chi}^0(\vec{q},\omega)$,
 are defined by
\be
\big[\chi^0(\vec{q},\omega)\big]_{ij}=-\frac{i}{\hbar\Omega}\int_0^\infty dt e^{i(\omega+i\delta)t}\langle 0\vert\big[A_i(t),A_j^\dagger(0)]\vert 0\rangle,
\ee
where
\bea
A_1(t) &=& e^{i(H_0+H_{SC})t/\hbar}\hat{O}(\vec{q})e^{-i(H_0+H_{SC})t/\hbar},\\
A_2(t) &=& e^{i(H_0+H_{SC})t/\hbar}(\Delta^d(-\vec{q}) - \Delta^{d\,\dagger}(\vec{q}))e^{-i(H_0+H_{SC})t/\hbar},\nn\\
A_3(t) &=& e^{i(H_0+H_{SC})t/\hbar}(\Delta^d(-\vec{q}) + \Delta^{d\,\dagger}(\vec{q}))e^{-i(H_0+H_{SC})t/\hbar},\nn
\eea
and $\hat{U}_{\vec{q}}$ is the effective interaction kernel,
\bea
\hat{U}_{\vec{q}} = \left(
\begin{array}{ccc}
V_{OO}(\vec{q})& 0 & 0\\
0& \frac{V_d(\vec{q})}{2}&0\\
0&0&\frac{V_d(\vec{q})}{2}
\end{array}
\right).
\eea

To assist with the analysis, we divide the $d$-wave competing pairing channel into phase ($A_2$) and amplitude ($A_3$) modes in order to enforce the time reversal symmetry in each of the 
of $A_i$ channels.   It will be shown later that because all of these three modes couple to each other due to the nature of the superconductivity and the space group symmetry, 
any theory ignoring these couplings is incomplete and might overlook important physics.
For the details, please see the Supplementary Material.

\section{Results}
From now on, we study the case of zero wavevector ($\vec{q}=0$) and neglect the subscript $\vec{q}$. In principle, our formalism can 
be applied for finite $\vec{q}$, but detailed knowledge in the $\vec{q}$-dependent interaction will be necessary in
order to obtain the correct dispersion of the resonance mode. For
proof of principle, an analysis of the 
zero-wavevector case suffices.
It is now clear, 
from the form of $\big[\chi^0(\vec{q},\omega)\big]_{ij}$, 
that the orbital ordering and the $d$-wave pairing channels are coupled because $\big[\chi^0\omega)\big]_{ij}$ is non-zero even for $i\neq j$. 
In Fig. \ref{fig:coupling}, we plot $\big[\chi^0(\omega)\big]_{12}$ and $\big[\chi^0(\omega)\big]_{13}$ for $\omega$ smaller than 
the particle-hole continuum edge $\Omega_{ph} \approx 2\Delta(k_F)$, which show the typical behavior for a Berry phase coupling: $\chi(\omega) \sim  \omega$\cite{leewcinsr2}.
A similar coupling occurs in the magnetic resonance mode of the cuprates in which the $\pi$-particle and spin-flip channels are coupled by 
symmetry\cite{demler1998,leewcinsr1,leewcinsr2}.

\begin{figure}
\includegraphics{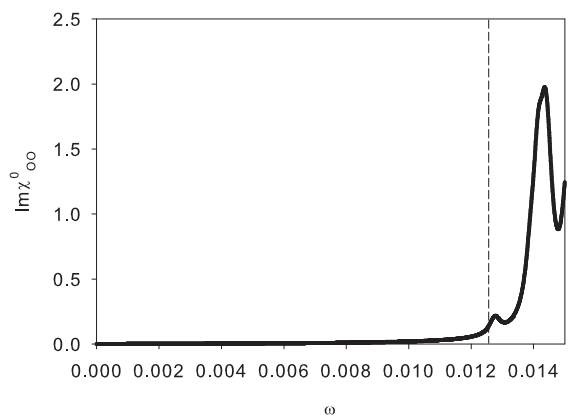}
\caption{\label{fig:chi0} The bare correlation function for the orbital-ordering channel. The magnitude of the gap is $\Delta_0 = 0.01$, and the edge of the particle-hole continuum
$\Omega_{ph}$ is indicated by the dashed line. To obtain a smoother
plot, we included a broadening factor $\delta=0.0002$ in our
computation of the imaginary part of the bare 
correlation function.}
\end{figure}

Since an $s_{\pm}$ superconductor is fully gapped on the Fermi surface, any of the two-particle correlation functions should have zero spectral weight below $\Omega_{ph}$ 
unless there exists a resonance mode with frequency $\omega_{res}$
which is a solution to 
\be
{\rm Det}(\left[\hat{\chi}(\omega_{res})\right]^{-1})=0.
\label{wres}
\ee
If $\Im \chi^0(\omega)$ has a sudden increase from zero at $\omega=\Omega_{ph}$, $\Re \chi^0(\omega)$ develops a weak logarithmic divergence at $\omega=\Omega_{ph}$ 
as can be shown from the Kramers-Kronig relations.  In this case,
Eq. (\ref{wres}) is satisfied if there is an effective {\it attractive} interaction in the orbital ordering channel, 
i.e., $V_{OO}(0) < 0$. 
In other words, an orbital resonance mode can generally arise if the fluctuations from the orbital-ordering transition
persist as the transition to the fully-gapped superconducting state obtains. 
This orbital resonance mode is an analogue of the magnetic resonance
mode in the cuprates.

\begin{figure}
\includegraphics{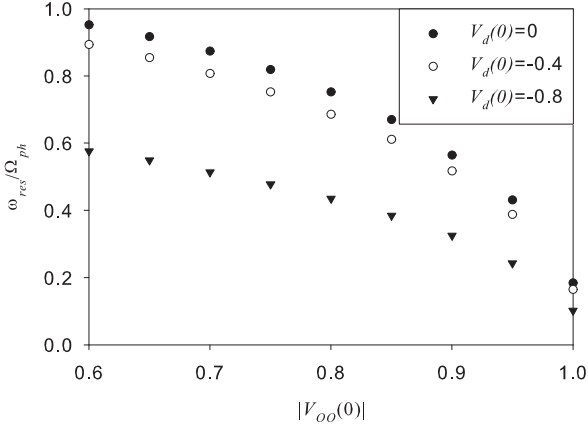}
\caption{\label{fig:wres} The ratio of the resonance frequency to the particle-hole continuum edge $\omega_{res}/\Omega_{ph}$ as a function of $\vert V_{OO}(0)\vert$ for 
different values of $V_d(0)$. At small interaction strength, $\omega_{res}/\Omega_{ph}\approx 1$. 
The instability of orbital ordering in the superconducting state occurs at $V_{OO}(0)\approx -1.2$, which has the solution of $\omega_{res}=0$ (not shown). 
Larger $\vert V_d(0)\vert$ tends to shift the resonance frequency to smaller values.
Since we are only interested in the superconducting state without orbital ordering, $0<\omega_{res}/\Omega_{ph}<1$.}
\end{figure}

In Fig. \ref{fig:chi0}, we plot the spectral weight of the bare
correlation function in the orbital-ordering channel. Clearly a jump from
zero to a finite value occurs at the edge of the 
particle-hole continuum. This signifies that $\omega_{res}$ always exists below $\Omega_{ph}$ for $V_{OO}(0) < 0$.
Let us discuss first the case without a residual $d$-wave pairing interaction $V_d(0)=0$.
Generally speaking, if the interaction strength $\vert V_{OO}(0)\vert$ is small, $\omega_{res}/\Omega_{ph}\approx 1$ and the spectral weight in this resonance mode 
is small. On the other hand, there exists a critical strength of $\vert V_{OO}(0)\vert$ for which Eq. \ref{wres} yields a solution of $\omega_{res}=0$ 
(in this calculation, this critical strength is $\vert V_{OO}(0)\vert\approx 1.2$). 
This marks the instability towards the orbital ordering transition. Since we are only interested in the superconducting state without orbital ordering, we have $0<\omega_{res}/\Omega_{ph}<1$.
This behavior is confirmed by our RPA calculations which are summarized in Fig. \ref{fig:wres}.

Note that the orbital resonance mode discussed so far will not be
present for a superconducting gap with nodes. 
The reason is that the orbital resonance mode emerges near zero wavevector. 
Consequently, the particle-hole continuum edge is roughly $2\Delta_{min}$, where $\Delta_{min}$ is the minimal gap on the Fermi surface. 
If $\Delta_{min} = 0$, which is the case of the gap symmetry with nodes on the Fermi surface, Eq. \ref{wres} can never be satisfied. 
Instead, a peak could still arise, but it will be damped or even completely washed out by the particle-hole continuum,  the precise details of which depend on the 
electronic structure of the material and thus is not symmetry protected.
The aforementioned case might be realized in cuprates in which nematic fluctuations (identical to the orbital fluctuations\cite{leewcnfl2012})
are argued to be present\cite{oganesyan2001,lawler2006} and the $d$-wave gap symmetry has nodes on the Fermi surface.
This is in a sharp contrast to the magnetic resonance mode which
usually appears near finite wavevector, for example, $(\pi,\pi)$ for
the cuprates and $(\pi,0)$ for iron pnictides.
In this case, the particle-hole edge, $\Omega_{ph}$, could be finite for both nodal and nodeless gap symmetry, and the existence of the magnetic resonance modes is determined primarily by the sign difference between $\Delta(\vec{k})$ and $\Delta(\vec{k}+\vec{q})$.

The Berry phase coupling gives rise to rich physics in the context of the resonance mode.
If only one of these two channels has strong fluctuations, the
resonance mode has an increased weight in the dominant channel.
Therefore, the $B_{1g}$ mode can be viewed as either an orbital
resonance mode or a $d$-wave excitonic pairing mode despite the
non-zero 
dynamical coupling between them.
If both channels, however, are equally strong, then the resonance mode has comparable weight in both channels. As a result,
the nature of the resonance mode becomes plasmonic\cite{leewcinsr1,leewcinsr2}, and the resonance frequency tends to be smaller as shown in Fig. \ref{fig:wres}.
It is interesting to point out that in iron pnictides, the relative
strength of the orbital ordering and the $d$-wave pairing fluctuations
could vary significantly from material to material.
This means that the nature of the resonance mode discussed above could be different for different families of the iron based superconductors.
In light of this observation, we argue that the interpretation of a $\delta$-like peak recently found in the Raman scattering measurement on Ba$_{0.6}$K$_{0.4}$Fe$_2$As$_2$ and 
Rb$_{0.8}$Fe$_{1.6}$Se$_2$ needs further consideration.

While Kretzschmar {\it et. al.}\cite{kretzschmar2012} have given a detailed fit to their data based on a model consisting of only competing pairing potentials, 
we would like to offer an alternative viewpoint based on the present theory. 
We note that from Fig. 3 in Ref.[\cite{kretzschmar2012}], all
of the peaks observed in Rb$_{0.8}$Fe$_{1.6}$Se$_2$ appear both above
and below the superconducting 
transition temperature $T_c$, while in Fig. 4(c), the peaks in the
$B_{1g}$ symmetry in Ba$_{0.6}$K$_{0.4}$Fe$_2$As$_2$ are present only
in the superconducting state.
This indicates that only Ba$_{0.6}$K$_{0.4}$Fe$_2$As$_2$ has truly superconductivity-induced collective modes of $B_{1g}$ symmetry. 
Since the orbital order operator, $\hat{O}(\vec{q}=0)$, defined in Eq. \ref{orderparameter} changes sign upon a rotation by $\pi/2$, it has $B_{1g}$ symmetry and consequently 
the $B_{1g}$ Raman spectroscopy is the ideal probe of the orbital resonance mode discussed in this paper. 
The first peak at $\omega=190 cm^{-1}$ is the feature of the particle-hole continuum edge, and we interpret the second peak at $\omega=140 cm^{-1}$ to be the 
orbital resonance mode.
For Rb$_{0.8}$Fe$_{1.6}$Se$_2$, because a structural phase transition is not observed and theoretical calculations also suggest an absence of ferro-orbital ordering
\cite{lv2011,luo2011,yin2012}, the orbital resonance mode should not
exist in this material. 

In principle, the orbital resonance mode can be detected either by probes in
the charge channel which are sensitive to B$_{1g}$ symmetry or
electron energy loss spectroscopy (EELS) because of its momentum resolution. EELS can directly measure the density-density correlation function $\chi(\vec{q},\omega)$ at finite momentum 
$\vec{q}$\cite{eels}. 
From a symmetry point of view, if $\vec{q}$ is parallel to either the
$\hat{x}$ or $\hat{y}$ directions, $\chi(\vec{q},\omega)$ will break the $C_4$ symmetry and thus it can couple to 
 orbital fluctuations. On the other hand, if $\vec{q}$ is along the
off-diagonal direction with respect to the $\hat{x}$ and $\hat{y}$ axes, $\chi(\vec{q},\omega)$ 
will still have $C_4$ symmetry so that it will not  couple to orbital fluctuations. As a result, for finite but small momentum $\vec{q}$, EELS can also reveal the 
orbital resonance mode discussed in this paper.


\section{Acknowledgement}
We thank E. Fradkin and A. J. Leggett for helpful discussions. We are also grateful to P. Abbamonte for a very useful discussion on EELS measurements.
This work is supported by the Center for Emergent 
Superconductivity, a DOE Energy Frontier Research Center, Grant 
No.~DE-AC0298CH1088.

\newpage
\section{appendix}

Here we provide all the details of the formalism used in the paper.
First, the tight-binding Hamiltonian $H_0$ is defined as
\bea
H_0 &=& \sum_{\vec{k},\sigma} \psi^\dagger_{\vec{k},\sigma}\left[\epsilon_+(\vec{k})\hat{I} + \epsilon_-(\vec{k})\hat{\tau}_3 + \epsilon_{xy}(\vec{k})\hat{\tau}_x\right]
\psi_{\vec{k},\sigma}\nn\\
&\equiv& \sum_{\vec{k},\sigma} \phi^\dagger_{\vec{k},\sigma}\hat{D}_{\vec{k},\sigma}\phi_{\vec{k},\sigma}
\eea
$\psi^\dagger_{\vec{k},\sigma} \equiv (d^\dagger_{\vec{k},xz,\sigma}, d^\dagger_{\vec{k},yz,\sigma})$, $\hat{D}_{\vec{k},\sigma} = 
diag. (E^{\alpha}_{\vec{k},\sigma},E^{\beta}_{\vec{k},\sigma})$, and
$\phi^\dagger_{\vec{k},\sigma} \equiv (\alpha^\dagger_{\vec{k},\sigma}, \beta^\dagger_{\vec{k},\sigma})$.
We introduce a unitary transformation $\hat{U}_{\vec{k},\sigma}$ such that 
\be
\hat{U}^\dagger_{\vec{k},\sigma}\left[\epsilon_+(\vec{k})\hat{I} + \epsilon_-(\vec{k})\hat{\tau}_3 + \epsilon_{xy}(\vec{k})\hat{\tau}_x\right]
\hat{U}_{\vec{k},\sigma} = \hat{D}_{\vec{k},\sigma},
\ee
and it can be derived that
\bea
\hat{U}_{\vec{k},\sigma} = \left(
\begin{array}{cc}
\cos\theta_k& \sin\theta_k\\
-\sin\theta_k& \cos\theta_k\\
\end{array}
\right),
\eea
where
\be
\cos 2\theta_k = \frac{\epsilon_-(\vec{k})}{H(\vec{k})}\,\,\,,\,\,\,\sin 2\theta_k = \frac{\epsilon_{xy}(\vec{k})}{H(\vec{k})}.
\ee
The eigenvalues are 
\bea
E^\alpha_{k\sigma}&\equiv& \epsilon_+(\vec{k}) + H(\vec{k})\,\,\,,\,\,\,E^\beta_{k\sigma}\equiv \epsilon_+(\vec{k}) - H(\vec{k}),\nn\\
\epsilon_+(\vec{k})&\equiv&\frac{\epsilon_{xz}(\vec{k}) + \epsilon_{yz}(\vec{k})}{2}, \nn\\
\epsilon_-(\vec{k})&\equiv&\frac{\epsilon_{xz}(\vec{k}) - \epsilon_{yz}(\vec{k})}{2}, \nn\\
H(\vec{k})&\equiv&\sqrt{\epsilon_-(\vec{k})^2 + \epsilon_{xy}^2},
\eea
and the corresponding eigenvectors can be expressed as 
$\psi_{\vec{k},\sigma} = \hat{U}_{\vec{k},\sigma} \phi_{\vec{k},\sigma}$.

The next step is to perform a Bogoliubov transformation for $H^\prime = H_0+H_{SC}$. 
We define $\Psi^\dagger_{\alpha,\vec{k}} \equiv (\alpha^\dagger_{\vec{k},\uparrow}, \alpha_{-\vec{k},\downarrow})$ and
$\Psi^\dagger_{\beta,\vec{k}} \equiv (\beta^\dagger_{\vec{k},\uparrow}, \beta_{-\vec{k},\downarrow})$.
Then we can rewrite
\bea
H_0 + H_{SC} &=& \sum_{\vec{k},\mu=\alpha,\beta} \Psi^\dagger_{\mu,\vec{k}}\left[ E^\mu_{\vec{k}} \hat{\tau}_3 + \Delta_\mu(\vec{k})\hat{\tau}_1\right]\Psi_{\mu,\vec{k}}\nn\\
&=&\sum_{\vec{k},\mu=\alpha,\beta} E^\mu_{SC}(\vec{k}) \Phi^\dagger_{\mu,\vec{k}}\hat{\tau}_3\Phi_{\mu,\vec{k}},
\eea
where $\Phi^\dagger_{\alpha,\vec{k}} \equiv (\alpha^{SC\,\dagger}_{+,\vec{k}}, \alpha^{SC\,\dagger}_{-,\vec{k}})$ and 
$\Phi^\dagger_{\beta,\vec{k}} \equiv (\beta^{SC\,\dagger}_{+,\vec{k}}, \beta^{SC\,\dagger}_{-,\vec{k}})$.

are the Bogoliubov quasiparticles in the $\alpha$ or $\beta$ bands.
The energies and the corresponding eigenvectors of the Bogoliubov quasiparticles are 
\bea
\Psi_{\mu,\vec{k}} &=& \hat{U}^{SC}_{\mu,\vec{k}} \Phi_{\mu,\vec{k}},\nn\\
E^\mu_{SC}(\vec{k}) &=& \sqrt{E^{\mu\,2}_{\vec{k}} + \Delta^2_\mu(\vec{k})}, 
\eea
where we have introduced another unitary transformation $\hat{U}^{SC}_{\mu,\vec{k}}$ such that
\bea
\hat{U}^{SC}_{\mu\vec{k}} = \left(
\begin{array}{cc}
\cos\omega_{\mu,\vec{k}}& \sin\omega_{\mu,\vec{k}}\\
-\sin\omega_{\mu,\vec{k}}& \cos\omega_{\mu,\vec{k}}\\
\end{array}
\right),
\eea
where
\be
\cos 2\omega_{\mu,\vec{k}} = \frac{E^\mu_{\vec{k}}}{E^\mu_{SC}(\vec{k})}\,\,\,,\,\,\,\sin 2\omega_{\mu,\vec{k}} = \frac{\Delta_\mu(\vec{k})}{E^\mu_{SC}(\vec{k})}.
\ee

All that is left is a series of long but straightforward caluclations. 
We express all the three channels $A_1, A_2, A_3$ in terms of the Bogoliubov quasiparticles $\{\alpha^{SC}_\pm,\beta^{SC}_\pm\}$, and then the 
bare correlation functions are

\begin{widetext}
\bea
\big[\hat{\chi}^0(\vec{q},\omega)\big]_{IJ}
&=& \frac{1}{\Omega}\sum_{\vec{k}} \big[
- \frac{A_I\,A_J}{\hbar\omega + i\delta
+ (E^\alpha_{SC}(k-q) + E^\alpha_{SC}(k))}
+ \frac{B_I\,B_J}{\hbar\omega + i\delta
- (E^\alpha_{SC}(k-q) + E^\alpha_{SC}(k))}
\big]\nn\\
&+& \big[
- \frac{C_I\,C_J}{\hbar\omega + i\delta
+ (E^\beta_{SC}(k-q) + E^\beta_{SC}(k))}
+ \frac{D_I\,D_J}{\hbar\omega + i\delta
- (E^\beta_{SC}(k-q) + E^\beta_{SC}(k))}
\big]\nn\\
&+& \big\{
E_I\,E_J\times\big[-\frac{1}{\hbar\omega + i\delta
+ (E^\alpha_{SC}(k-q) + E^\beta_{SC}(k))} + \frac{1}{\hbar\omega + i\delta
- (E^\alpha_{SC}(k-q) + E^\beta_{SC}(k))}\big]\nn\\
&+& F_I\,F_J\times\big[
 \frac{1}{\hbar\omega + i\delta
- (E^\beta_{SC}(k-q) + E^\alpha_{SC}(k))} - \frac{1}{\hbar\omega + i\delta
+ (E^\beta_{SC}(k-q) + E^\alpha_{SC}(k))}\big]
\big\}
\eea
\end{widetext}
where
\bea
A_1 &=& B_1= \cos(\theta_{k-q}+\theta_k) \times \sin\left(\omega_{\alpha,\vec{k}}+\omega_{\alpha,\vec{k}-\vec{q}}\right)\nn\\
A_2 &=& -B_2 = -d_{\vec{k}}\times \cos\left(\omega_{\alpha,\vec{k}}+\omega_{\alpha,\vec{k}-\vec{q}}\right)\nn\\
A_3 &=& -B_3 = -d_{\vec{k}}\times \cos\left(\omega_{\alpha,\vec{k}}-\omega_{\alpha,\vec{k}-\vec{q}}\right)\nn\\
C_1 &=& D_1= -\cos(\theta_{k-q}+\theta_k) \times \sin\left(\omega_{\beta,\vec{k}}+\omega_{\beta,\vec{k}-\vec{q}}\right)\nn\\
C_2 &=& -D_2 = -d_{\vec{k}}\times \cos\left(\omega_{\beta,\vec{k}}+\omega_{\beta,\vec{k}-\vec{q}}\right)\nn\\
C_3 &=& -D_3 = -d_{\vec{k}}\times \cos\left(\omega_{\beta,\vec{k}}-\omega_{\beta,\vec{k}-\vec{q}}\right)\nn\\
E_1 &=& \sin(\theta_{k-q}+\theta_k)\times \sin\left(\omega_{\beta,\vec{k}}+\omega_{\alpha,\vec{k}-\vec{q}}\right)\nn\\
E_2 &=& E_3=0\nn\\
F_1 &=& \sin(\theta_{k-q}+\theta_k)\times \sin\left(\omega_{\alpha,\vec{k}}+\omega_{\beta,\vec{k}-\vec{q}}\right)\nn\\
F_2 &=& F_3=0\nn.\\
\eea

\end{document}